\documentclass[12pt,thmsa]{article}

\begin{document}

\author{A. Yu. Zyuzin}
\date{A.F.Ioffe Physical-Technical Institute 194021 Saint-Petersburg, Russia \\
and\\
Max-Planck-Institut f\"{u}r Festk\"{o}rperforschung BP166, F-38042 Grenoble,
France}
\title{Superfluorescence of Photonic Paint }
\date{A.F.Ioffe Physical-Technical Institute 194021 Saint-Petersburg, Russia \\
and\\
Max-Planck-Institut f\"{u}r Festk\"{o}rperforschung BP166, F-38042 Grenoble,
France}
\maketitle

\begin{abstract}
We consider the cooperative decay of incoherently pumped atoms in a
disordered medium, where light undergoes multiple scattering. It is shown
that the cooperation number, which determines the duration and amplitude of
superfluorescent impulses, is given by the number of atoms along a diffusive
trajectory of the light propagating through the medium. We also consider the
problem of reflection of a probe wave during cooperative emission.
\end{abstract}

\section{Introduction\protect\vspace{0cm}}

There is growing interest active photonic paints. These are media in which
light undergoes multiple random scattering, resulting in a diffusive
propagation of radiation, while interacting with atoms that can be pumped to
obtain a positive population difference. The reflection and transmission of
the electromagnetic waves through such a cavity has been extensively studied
over the past decade. The speckle pattern resulting from scattering has an
average enhancement in the direction opposite the direction of the incident
radiation\cite{Kuga84}, (a comprehensive review of other statistical
properties of the speckle of reflected and transmitted waves is given in\cite
{Genack90}).

Feedback provided by scattering in such a random cavity can serve to set up
laser oscillations~\cite{Letokhov67}. The laser action in a powdered laser
materials \cite{Ter-Gabrielyan91}\cite{Gouedard93}, laser dye solutions with
scattering nanoparticles~\cite{Lawandy94}, and dye-doped microdroplets
containing Intralipid as a scatterer~\cite{Taniguchi96} has recently been
reported. These experiments concentrated mostly on temporal and spatial
properties of emission.

Recently, the proposed~\cite{Zyuzin94} enhancement of the weak localization
peak in backscattering from an amplifying photonic paint was observed~\cite
{Wiersma95}.

The relevant question concerning recent observations of generation of light
in active photonic paints~\cite{Ter-Gabrielyan91} \cite{Gouedard93}is to
what extent this phenomenon is reminiscent of superfluorescence~\cite
{Dicke54} \cite{Bonifatio75}(i.e., the cooperative decay of an incoherently
pumped system of dipole transitions, started by initial noise or an external
electromagnetic field), which has usually been studied in systems without
scattering.

Here we consider the cooperative decay of incoherently pumped atoms in a
random cavity, which is a slab of thickness $L$($L\gg l$, where $l$ is the
mean free path of radiation). This geometry is often used in experiments.
The time that light spends in this cavity is of order $\frac{L^2}D$, where $%
D $ is diffusion constant. This time is to be compared with the energy
exchange time between atoms and field. We show that if the latter is greater
than $\frac{L^2}D$, then after some delay, the system will generate a
superfluorescent pulse of hyperbolic secant form.

The duration of the superfluorescent pulse is $\tau _{rad}N_C^{-1}$~\cite
{Dicke54} , where $\tau _{rad}$ is the time of radiative decay of a single
atom and $N_C$ is the cooperation number, i.e., the number of atoms that
take part in cooperative decay. We find that in disordered systems, this
number is $N_C\sim \rho \frac{\lambda ^2L^2}l$ ( $\lambda $ is the
wavelength of the radiation and $\rho $ is the density of active atoms, such
that $\rho \lambda ^3\gg 1$), i.e., it is equal to the number of atoms in a
tube with cross section $\lambda ^2$ and length of the order of $\frac{L^2}l$
, which is the length of a diffusive trajectory of radiation.

The intensity of radiation of cooperating atoms at the maximum of the
superfluorescent pulse is $\sim \frac{N_C^2}{\tau _{rad}}$~\cite{Dicke54}.
We show that the diffusive slab radiates at maximum as a system of $\frac
V{N_C}$ independent groups of cooperating atoms, and at the peak of
superfluorescent pulse, the intensity emitted by the slab is $\sim \frac{%
N_C^2}{\tau _{rad}}\times \frac V{N_C}$. $V$ is the volume of the slab.

The maximum cooperation number for given $\tau _{rad}$ and density of active
atoms is determined by the condition that the time of energy exchange
between atoms and the field equals the time that light spends in the cavity.
From this condition, we find that the maximum cooperation number in the
random cavity is $N_C^{max}\sim \lambda \sqrt{c\rho \tau _{rad}}$, where $c$
is the speed of radiation in the slab.

These results are valid in the case of weak dephasing processes and long
relaxation of population difference. Below we take into account the effect
of dephasing on superfluorescence.

In the limit of the large escape time $\frac{L^2}D$ of radiation, atoms
exchange energy with the field many times, so stimulated emission becomes
important and the system exhibits oscillatory behavior.

We also consider the reflection of the probe wave during decay of the pumped
system.

\section{Basic equations.}

We model a random medium in the following way. The dielectric function $%
\epsilon \left( \overrightarrow{r}\right) $ of the medium, which contains
active atoms, is a random function of position, such that $\left\langle
\epsilon \left( \overrightarrow{r}\right) \right\rangle $. Scattering of
light is due to fluctuations of the dielectric function with white-noise
like variance $\left\langle \delta \epsilon \left( \overrightarrow{r}\right)
\delta \epsilon \left( \overrightarrow{r^{\prime }}\right) \right\rangle =%
\frac{\lambda ^4}{4\pi ^3l}\delta \left( \overrightarrow{r}-\overrightarrow{%
r^{\prime }}\right) $.

We consider the case of a weakly disordered system $l\gg \lambda $, with
dimensions larger the than mean free path, so propagation of the field can
be described as a diffusion process with diffusion constant $D=\frac{cl}3$, $%
c$ is the speed of light in the medium.

The coupling between the polarization density $\frac 12\left\{ e^{i\omega
t}P\left( \overrightarrow{r};t\right) +e^{-i\omega t}P^{*}\left( 
\overrightarrow{r};t\right) \right\} $, averaged over scales smaller than $%
\lambda $, the population difference density $\Delta N\left( \overrightarrow{%
r};t\right) $, and the field $\frac 12\left\{ e^{i\omega t}E\left( 
\overrightarrow{r};t\right) +e^{-i\omega t}E^{*}\left( \overrightarrow{r}%
;t\right) \right\} $ can be described by the classical Maxwell-Bloch
equations. In this approach, amplified spontaneous emission noise is
neglected, which is a good approximation for superfluorescence~\cite
{Bonifacio71}. $P\left( \overrightarrow{r};t\right) $and $E\left( 
\overrightarrow{r};t\right) $ are slowly time-varying complex quantities,
which we consider to be scalars; $\omega $ is the atomic frequency.

First two Maxwell-Bloch equations have the form~\cite{Siegman86}.

\begin{equation}
\left[ \frac d{dt}+\gamma \right] P\left( \overrightarrow{r};t\right) =\frac{%
i\left| \mu \right| ^2}\hbar \Delta N\left( \overrightarrow{r};t\right)
E\left( \overrightarrow{r};t\right)  \label{eq1}
\end{equation}

\begin{equation}
\frac d{dt}\Delta N\left( \overrightarrow{r};t\right) =-\frac i{2\hbar
}\left\{ P^{*}\left( \overrightarrow{r};t\right) E\left( \overrightarrow{r}%
;t\right) -P\left( \overrightarrow{r};t\right) E*\left( \overrightarrow{r}%
;t\right) \right\}  \label{eq2}
\end{equation}

Here $\gamma $ is the inverse dephasing time and $\mu $ is the electric
dipole moment.

\strut It is assumed that the population inversion relaxation time is longer
than the delay time of the superfuorescent pulse. We also neglect
inhomogeneous broadening.

The quantities $\Delta N\left( \overrightarrow{r};t\right) $ and $\frac{%
P\left( \overrightarrow{r};t\right) }\mu $ are components of the local Bloch
vector averaged over scales smaller than the wavelength of the radiation.
The rate at which its length decreases, according to (1) and (2), is
determined by $\gamma ^{-1}$.

The field wave equation for the slow time-varying component $E\left( 
\overrightarrow{r};t\right) $ has the form

\begin{equation}
i\frac{dE\left( \overrightarrow{r};t\right) }{dt}-\left\{ -\frac{c^2}{%
2\omega }\Delta -\frac{\omega \epsilon \left( \overrightarrow{r}\right) }%
2\right\} E\left( \overrightarrow{r};t\right) =2\pi \omega P\left( 
\overrightarrow{r};t\right)  \label{eq3}
\end{equation}

Although $E\left( \overrightarrow{r};t\right) $ and $P\left( \overrightarrow{%
r};t\right) $ vary slowly in time, they still contain spatial random phases,
which result from random interference between waves coming to the point $%
\overrightarrow{r}$ via different diffusive trajectories. To get rid of
these phase factors, it is convenient to consider the diffusion propagator $%
D\left( \overrightarrow{r};t_1,t_2\right) $, which determines the
correlation function of the polarization density and field :

\begin{equation}
\left\langle E\left( \overrightarrow{r};t_1\right) E^{*}\left( 
\overrightarrow{r};t_2\right) \right\rangle =4\pi k^3\omega D\left( 
\overrightarrow{r};t_1,t_2\right)  \label{eq4}
\end{equation}

Correlation functions involving the polarization density can be obtained by
using Eq.(1).

To obtain the equation for the diffusion propagator, it is convenient to
eliminate the polarization density from Eqs.(1) and (3). Then the usual
diagram technique ~\cite{Abrikosov63} makes it possible to calculate the
average of the product $E\left( \overrightarrow{r};t_1\right) E^{*}\left( 
\overrightarrow{r};t_2\right) $ .

Considering the evolution of the Bloch vector from time $t=0$, at which a
positive population difference is created, we obtain for the diffusion
propagator

\[
\left\{ \frac d{dt_1}+\frac d{dt_2}-D\overrightarrow{\nabla }^2\right\}
D\left( \overrightarrow{r};t_1,t_2\right) =f\left( \overrightarrow{r}%
;t_1,t_2\right) + 
\]

\begin{equation}
+\frac 1{\rho \tau _0^2}\int\limits_0^{t_1}dt\exp \left\{ -\gamma \left(
t_1-t\right) \right\} \Delta N\left( \overrightarrow{r},t\right) D\left( 
\overrightarrow{r};t,t_2\right) +  \label{eq5}
\end{equation}

\[
+\frac 1{\rho \tau _0^2}\int\limits_0^{t_2}dt\exp \left\{ -\gamma \left(
t_2-t\right) \right\} \Delta N\left( \overrightarrow{r},t\right) D\left( 
\overrightarrow{r};t_1,t_2\right) 
\]

Here $\tau _0=\sqrt{\frac \hbar {2\pi \rho \omega \left| \mu \right| ^2}}=%
\sqrt{\frac{4\pi \tau _{rad}}{3\omega \rho \lambda ^3}}$ is the
characteristic time of energy exchange between the field and the atomic
system~\cite{Bonifacio75}, $\rho $ is the density of active atoms, and $\tau
_{rad}^{-1}=\frac{8\pi ^2\left| \mu \right| ^2}{3\hbar \lambda ^3}$ is the
radiative decay time of a single atom.

The function $f\left( \overrightarrow{r};t_1,t_2\right) $ depends on initial
conditions. Here we choose the initial condition such that $\left\langle
P\left( \overrightarrow{r},t=0\right) \right\rangle =0$ and $\left\langle
P\left( \overrightarrow{r},t=0\right) P^{*}\left( \overrightarrow{r^{\prime }%
},t=0\right) \right\rangle =\rho \left| \mu \right| ^2\delta \left( 
\overrightarrow{r}-\overrightarrow{r^{\prime }}\right) $. This initial
condition corresponds to an initial incoherent state. In this case

\begin{equation}
f\left( \overrightarrow{r};t_1,t_2\right) =\rho \left| \mu \right| ^2\exp
\left\{ -\gamma \left( t_1+t_2\right) \right\}  \label{eq6}
\end{equation}
for times greater than the mean free time of radiation $\frac lc$.

The equation for the mean population inversion density can be obtained by
using Eqs. (1), (2) and (5):

\begin{equation}
\frac{d\Delta N\left( \overrightarrow{r},t\right) }{dt}=-\frac{\left( 2\pi
\right) ^3}{\hbar \rho \lambda ^3\tau _0^2}\int\limits_0^tdt\exp \left\{
-\gamma \left( t-t^{\prime }\right) \right\} \Delta N\left( \overrightarrow{r%
},t^{\prime }\right) \left\{ D\left( \overrightarrow{r};t^{\prime },t\right)
+D\left( \overrightarrow{r};t,t^{\prime }\right) \right\}
\end{equation}

For the population difference we choose $\Delta N\left( \overrightarrow{r}%
,t=0\right) =\Delta N>0$ as the initial condition ($\Delta N=\rho $).

The usual boundary conditions for the diffusion propagator are $D\left( 
\overrightarrow{r};t_1,t_2\right) =0$ on an open surface and $%
\overrightarrow{n}\overrightarrow{\nabla }D\left( \overrightarrow{r}%
;t_1,t_2\right) =0$ on a reflecting surface; $\overrightarrow{n}$ is normal
to the reflecting surface.

The diffusion approach is justified if the time of energy exchange between
atoms and field is greater than mean the free time of radiation, $\tau _0\gg
\frac lc$.

\section{Cooperative decay in photonic paint}

Below we consider a slab of thickness $L$ ($L\gg l$). Let $z$ be the
coordinate across the slab $L\succeq z\succeq 0$. It is convenient to study
the solution of Eq. (5) in the form

\begin{equation}
D\left( \overrightarrow{r};t_1,t_2\right) =\sqrt{\frac L2}\sum_{n=1}^\infty
\Psi _n\left( z\right) D_n\left( t_1,t_2\right)
\end{equation}
where $\Psi _n\left( z\right) =\sqrt{\frac 2L}\sin \frac{\pi nz}L$ is an
eigenfunction of the diffusion equation with boundary condition $\Psi
_n\left( z\right) =0$ at the free boundaries $z=0,L$.

Let us consider the initial evolution of the diffusion propagator, when the
population difference does not depend on time. For the coefficients in (8),
we obtain from (5)

\begin{eqnarray}
\left\{ \frac d{dt_1}+\frac d{dt_2}+\omega _n\right\} D_n\left(
t_1,t_2\right) &=&\sqrt{\frac 2L}\rho \left| \mu \right| ^2\int dz\Psi
_n\left( z\right) \exp \left\{ -\gamma \left( t_1+t_2\right) \right\} + 
\nonumber \\
&&+\frac{\Delta N}{\rho \tau _0^2}\int\limits_0^{t_1}dt\exp \left\{ -\gamma
\left( t_1-t\right) \right\} D_n\left( t,t_2\right) + \\
&&+\frac{\Delta N}{\rho \tau _0^2}\int\limits_0^{t_2}dt\exp \left\{ -\gamma
\left( t_2-t\right) \right\} D_n\left( t_1,t_2\right)  \nonumber
\end{eqnarray}

Here $\omega _n=\frac{D\pi ^2n^2}{L^2}$ is an eigenvalue of the diffusion
equation. Solving Eq. (9) via the Laplace transform with initial conditions $%
D_n\left( 0,t_2\right) =D_n\left( t_1,0\right) =0$ (the field vanishes at $%
t=0$), we obtain

\begin{equation}
D_n\left( t,t\right) \sim \exp \left\{ \left( \sqrt{\left( \frac{\omega _n}%
2-\gamma \right) ^2+\frac{4\Delta N}{\rho \tau _0^2}}-\frac{\omega _n}%
2-\gamma \right) t\right\}
\end{equation}

The critical value of positive inversion density $\Delta N_n$, above which
the growth rate of a particular diffusion mode $Z_n=\sqrt{\left( \frac{%
\omega _n}2-\gamma \right) ^2+\frac{4\Delta N}{\rho \tau _0^2}}-\frac{\omega
_n}2-\gamma $ becomes positive, is $\Delta N_n=\frac{\omega _n\gamma }2\rho
\tau _0^2$~\cite{Letokhov68}. More detailed calculations of (10) are given
in the next section.

To proceed further in solving Eqs. (5) and (7), we make two approximations.

1). Below we consider the case of fast escape of radiation from the system,
where $\omega _1\gg \frac d{dt_1},\frac d{dt_2},\gamma $ (or, according to
(10), $\tau _0\omega _1>1$ for weak dephasing), so we can neglect the time
derivative in Eq. (5). In the language of superfluorescence, this situation
corresponds to the case in which there is no energy exchange between the
emitted field and atomic subsystem~\cite{Bonifacio71}. The field serves only
to develop correlation between atoms.

2). We consider only the most unstable mode $D_1\left( t_1,t_2\right) $. At $%
t=0$, the off-diagonal elements of $\Delta N_{nm}\equiv \int dz\Delta
N\left( z,t\right) \Psi _n\left( z\right) \Psi _m\left( z\right) $ are zero
by definition, and interaction between modes is irrelevant for most of the
time of decay. We therefore assume that the interaction of the first
diffusion mode $D_1\left( t_1,t_2\right) $ with higher modes does not
qualitatively change the description of cooperative decay.

Under these assumptions the equation for the diffusion propagator has the
form

\begin{eqnarray}
D_1\left( t_1,t_2\right) &=&\frac 4{\pi \omega _1}\rho \left| \mu \right|
^2\exp \left\{ -\gamma \left( t_1+t_2\right) \right\} +  \nonumber \\
&&+\frac 1{\rho \omega _1\tau _0^2}\int\limits_0^{t_1}dt\exp \left\{ -\gamma
\left( t_1-t\right) \right\} \Delta N_{11}\left( t\right) D_1\left(
t,t_2\right) + \\
&&+\frac 1{\rho \omega _1\tau _0^2}\int\limits_0^{t_2}dt\exp \left\{ -\gamma
\left( t_2-t\right) \right\} \Delta N_{11}\left( t\right) D_1\left(
t_1,t_2\right)  \nonumber
\end{eqnarray}

and for the population difference

\begin{equation}
\frac d{dt}\Delta N_{11}\left( t\right) =-\frac{k^3}\hbar \left\{ \frac
8{3\pi }\omega _1D_1\left( t,t\right) -\rho \left| \mu \right| ^2\exp
\left\{ -2\gamma t\right\} \right\}
\end{equation}

Introducing

\begin{equation}
D_1\left( t_1,t_2\right) \equiv \exp \left\{ -\gamma \left( t_1+t_2\right)
\right\} G\left( \chi \left( t_1\right) ;\chi \left( t_2\right) \right)
\end{equation}

in (11),where

\begin{equation}
\chi \left( t\right) =\frac 1{\rho \omega _1\tau
_0^2}\int\limits_0^tdt\Delta N_{11}\left( t\right)
\end{equation}

we obtain

\begin{equation}
G\left( \chi _1;\chi _2\right) =\frac 4{\pi \omega _1}\rho \left| \mu
\right| ^2+\int\limits_0^{\chi _1}d\chi G\left( \chi ;\chi _2\right)
+\int\limits_0^{\chi _2}d\chi G\left( \chi _1;\chi \right)
\end{equation}

Equation (15) can be solved by a Laplace transform as

\begin{equation}
G\left( \chi _1;\chi _2\right) =\frac{4\rho \left| \mu \right| ^2}{\pi
\omega _1}\int\limits_{C-i\infty }^{C+i\infty }\frac{dz_1}{2\pi i}\frac{dz_2%
}{2\pi i}\frac{\exp \left\{ z_1\chi _1+z_2\chi _2\right\} }{z_1z_2-z_1-z_2}
\end{equation}

The asymptotic form of (16) for $\chi _1=\chi _2\equiv \chi >1$ is

\begin{equation}
G\left( \chi ;\chi \right) \simeq \frac{2\rho \left| \mu \right| ^2}{\pi
\omega _1}\frac{\exp \left\{ 4\chi \right\} }{\sqrt{\pi \chi }}
\end{equation}

The equation for the population difference ($\chi >1$) is

\begin{equation}
\frac{d^2}{dt^2}\chi =-\frac{8k^3}{3\pi \hbar \rho \tau _0^2}G\left( \chi
;\chi \right) \exp \left\{ -2\gamma t\right\}
\end{equation}

Taking into account only exponential factors, we obtain the solution of Eq.
(18) :

\begin{equation}
\Delta N_{11}\left( t\right) =\delta N\tanh \left\{ \frac{2\delta N\left(
t_0-t\right) }{\rho \tau _0^2\omega _1}\right\} +\frac{\gamma \rho \tau
_0^2\omega _1}2
\end{equation}

Here we introduce $\delta N=\Delta N-\frac{\gamma \rho \tau _0^2\omega _1}2$%
; $\Delta N$ is the population difference at the beginning of exponential
growth of radiative intensity, when deviation from the initial population
difference is small ($\Delta N=\rho $).

The delay time in (19) is $t_0=\frac{\rho \tau _0^2\omega _1}{2\delta N}\ln
\left\{ \frac{\delta N}\rho \sqrt{\rho \lambda ^3}\frac L{\sqrt{l\lambda }%
}\right\} $. In deriving this expression we took into account the relation
between the time of energy exchange between atoms and field and $\left| \mu
\right| $ , which enters into the initial condition for polarization density.

The radiative intensity is proportional to $\frac{d\Delta N_{11}}{dt}$, and
is emitted as a hyperbolic-secant pulse. The result (19) coincides with that
of the Markov theory of superfluorescence in a system without scattering~%
\cite{Dicke54,Bonifacio75}. The difference is in definition of the
cooperation number.

It follows from Eq. (19) that in the case of weak dephasing, the duration of
a superfluorescent pulse is $\frac{\tau _0^2\omega _1}4\equiv \tau
_{rad}N_C^{-1}$, where $\tau _{rad}$ is the time of radiative decay of a
single atom and $N_C=\frac{18\rho \lambda ^2L^2}{\pi ^2l}$ is the
cooperation number, i.e., the number of atoms that take part in the
cooperative decay. This is equal to the number of atoms in a tube of cross
section $\lambda ^2$ with the length of the diffusive trajectory $\frac{L^2}%
l $. The maximum of the cooperation number is determined by the condition $%
\tau _0\omega _1=1$, whereupon $N_C^{max}=2\lambda \sqrt{6c\rho \tau _{rad}}$%
. Under this condition, atoms can exchange energy with the field only once,
i.e., stimulated emission can be neglected. We note that for a given density 
$\rho $, decay time $\tau _{rad}$, and velocity, the maximum cooperation
number in a disordered system is smaller, than in a pencil-shaped system
without scattering~\cite{Arecchi70}.

The maximum emitted radiation is $V\frac d{dt}\Delta N_{11}\left(
t=t_0\right) $ ( $V$ is the volume of the slab). It can also be written N$%
\times \frac{N_C^2}{\tau _{rad}}$, where N$=\frac{V\rho }{N_C}$ is the
number of cooperative regions in the slab. The cooperative decay in a
diffusive medium can therefore be interpreted as the independent cooperative
decay of N$=\frac{V\rho }{N_C}$ systems, each consisting of $N_C$ atoms.

Dephasing processes increase the duration of a pulse by the factor $\frac
\rho {\Delta N-\frac{\gamma \rho \tau _0^2\omega _1}2}$, and decrease the
peak intensity by the square of this factor. Note that this result coincides
with that for a system without scattering~\cite{Andreev77 }.

If $1\gg \tau _0\omega _1$, atoms exchange energy with the field many times.
In this case we expect spiking of intensity. The frequency of spiking can be
estimated~\cite{Siegman86} from Eq. (10) as $\sqrt{\left| \frac{4\Delta N}{%
\rho \tau _0^2}\right| -\left( \frac{\omega _1}2-\gamma \right) ^2}$. To
obtain this expression we insert a negative value of the population
inversion~\cite{Siegman86} into (10) (this situation will occur after the
pumped atoms exchange energy with the field).

\section{Amplification in the backward direction}

Correlation between pumped atoms can also be due to the external field,
which stimulates emission in the forward direction in a system without
scattering~\cite{Siegman86}. In a disordered system one might expect
enhancement of emission in the backward direction.

Here we consider the reflection of a weak probe plane wave with frequency
during the development of superfluorescent emission. The amplitude of the
probe is low, so the effect of the external field on emission can be
neglected. We can also neglect interference between the external field with
the emitted one, because the initial state of polarization is incoherent.
Below we consider in detail the linear stage of decay when the inversion
density is high enough to produce only the lowest diffusion mode
instability, $\Delta N=\Delta N_1\left( 1+\delta \right) $ , $\delta \ll 1$.
This situation resembles the experimental setup of~\cite{Wiersma95}.

It is convenient to calculate the albedo, which is the ratio between the
intensities of the reflected and incident fields. The time-dependent albedo
can be expressed as~\cite{Akkermans86}

\begin{equation}
\alpha \left( \overrightarrow{q};t\right) =\frac c{4\pi
l^2}\int\limits_0^\infty dzdz^{\prime }\exp \left\{ -\frac{z+z^{\prime }}%
l\right\} \int d\overrightarrow{\rho }\left\{ 1+\cos \overrightarrow{q}%
\overrightarrow{\rho }\right\} D\left( z,z^{\prime },\overrightarrow{\rho }%
;t,t\right)
\end{equation}
Here $\overrightarrow{q}$ is the sum of the incident and outgoing wave
vectors, and $\overrightarrow{\rho }$ is the position in the plane.
Diffusion propagator (20) obeys Eq. (5) with the substitution of $\delta
\left( \overrightarrow{r}-\overrightarrow{r^{\prime }}\right) $ for $f\left( 
\overrightarrow{r};t_1,t_2\right) $. We also assume that the incident wave
is close to the normal to the surface.

The first term describes diffusion scattering, and the second term describes
the interference part, which is strongly peaked in the backward direction.
The physical mechanism of the interference contribution is exhaustively
discussed in the literature; see, for example Refs.. ~\cite{Kuga84,Zyuzin94}
and references therein.

The diffusion propagator can be represented as

\begin{equation}
D\left( \overrightarrow{r},\overrightarrow{r^{\prime }};t_1,t_2\right)
=\sum\limits_n\Psi _n\left( z\right) \Psi _n\left( z^{\prime }\right) \exp
\left\{ i\overrightarrow{q}\left( \overrightarrow{\rho }-\overrightarrow{%
\rho ^{\prime }}\right) \right\} \stackrel{\symbol{94}}{D}_n\left(
q;t_1,t_2\right)
\end{equation}

The Laplace transform of Eq. (5) for time-independent $\Delta N>0$ yields

\begin{equation}
\stackrel{\symbol{94}}{D}_n\left( q;t_1,t_2\right) =\int\limits_{-i\infty
+C}^{i\infty +C}\frac{dp_1dp_2}{\left( 2\pi i\right) ^2}\frac{\exp \left(
p_1t_1+p_2t_2\right) }{p_1p_2\left\{ p_1+p_2+\Omega _n\left( q\right) -\frac{%
\Delta N}{\rho \tau _0^2}\left( \frac 1{p_1+\gamma }+\frac 1{p_2+\gamma
}\right) \right\} }
\end{equation}

where the real part of the integration contour passes to the left of all
singularities, and $\Omega _n\left( q\right) =Dq^2+\omega _n$ is the
eigenvalue of the diffusion equation for the slab geometry.

At $t_1=t_2$, integrating over the difference $p_1-p_2$ in (22), we obtain
for the first mode at $t>\omega _1^{-1}$

\begin{equation}
\stackrel{\symbol{94}}{D}_n\left( q;t,t\right) =\frac{\left( 2\gamma \right)
^{\frac 32}}{\sqrt{\omega _1\left( \omega _1+2\gamma \right) }}%
\int\limits_{-i\infty +C}^{i\infty +C}\frac{dp}{2\pi i}\frac{\exp pt}{p\sqrt{%
p-Z_1\left( q\right) }\left( p+\sqrt{\frac{\gamma \left( \omega _1+2\gamma
\right) }{2\omega _1}}\sqrt{p-Z_1\left( q\right) }\right) }
\end{equation}

Here we introduce $Z_1\left( q\right) =\frac{2\gamma \omega _1}{\omega
_1+2\gamma }\left( \delta -\frac{Dq^2}{\omega _1}\right) $, which is the
growth rate of $\stackrel{\symbol{94}}{D}_n\left( q;t,t\right) $.

For moderate times $\frac{t\omega _1Z_1^2}{\gamma \left( \omega _1+2\gamma
\right) }<1$, we obtain from (23)

\begin{equation}
\stackrel{\symbol{94}}{D}_n\left( q;t,t\right) =\frac{2\left( \exp \left\{
Z_1\left( q\right) t\right\} -1\right) }{\omega _1\left\{ \delta -\frac{Dq^2%
}{\omega _1}\right\} }
\end{equation}

This expression is valid for either sign of $Z_1$, i.e., above as well as
below threshold.

Taking into account that $\Psi _1\left( z\right) =\sqrt{\frac 2L}\sin \frac{%
\pi z}L$ , we obtain the singular contribution to the albedo from the first
mode :

\begin{equation}
\delta \alpha \left( q,t\right) =\frac{3l}{\pi L}\left\{ \frac{\exp \left\{
Z_1\left( 0\right) t\right\} -1}\delta +\frac{\exp \left\{ Z_1\left(
q\right) t\right\} -1}{\left\{ \delta -\frac{Dq^2}{\omega _1}\right\} }%
\right\}
\end{equation}

Below threshold the albedo is saturated. The peak at large times has a
laplacian form $\propto \frac 1{\left| \delta \right| +\frac{Dq^2}{\omega _1}%
}$. At threshold and above there is narrowing of the peak with increasing
time. Exactly at threshold the albedo is linear with time, and above
threshold the albedo grows exponentially.

\section{Conclusions}

To summarize, superfluorescent emission of active photonic paint develops
due to the cooperation of atoms along a diffusive trajectory through a
system with cross-sectional dimensions of the order of a wavelength. The
pulse therefore becomes narrower with decreasing mean free path of radiation
until the cooperation number reaches its maximum value. The maximum
cooperation number does not depend on disorder.

An external field enhances emission in backward direction. The peak sharpens
in coherent backscattering during cooperative decay in a disordered system.

We thank A.V. Gol,tsev for useful suggestions. This work was supported by
the Russian Fund for Fundamental Research under Grant number 97-02-18078.

\end{document}